\documentclass[12pt]{iopart}
\usepackage{graphicx}
\begin{document}
\title{Heavy Quark Production from Relativistic Heavy Ion Collisions}

\title[Heavy Quark Production.....]{}

\author{Mohammed Younus and Dinesh K. Srivastava}

\address{Variable Energy Cyclotron Center, 1/AF, Bidhannagar,
 Kolkata 700 064, India}

\date{\today}

\begin{abstract}

We study the production of heavy quarks, charm at BNL-RHIC
($\sqrt{s}$ =  200 GeV/nucleon) and CERN-LHC
($\sqrt{s}$ =  5.5 TeV/nucleon) and bottom at CERN-LHC from
 heavy ions colliding at relativistic energies. We consider
initial fusion of gluons (and quark- anti-quark annihilation),
pre-thermal parton interactions and interactions
in thermalized quark gluon plasma.
We also consider free-streaming partons as another extreme
and compare the results with those from a thermalized plasma of partons.
The pre-thermal contribution is calculated by considering interaction
among partons having large transverse momenta (jet-partons) after the initial
interaction, and from passage of these partons through a thermalized
quark gluon plasma. Charm production from pre-thermal processes is found
to be comparable to that from prompt (initial) interactions at LHC.
 It is suggested that this may have important implications for the study of
nuclear modification factor,
R$_{\rm {AA}}$ as well as for back-to-back
correlation of heavy quarks and production of dileptons
 having a large mass.

\end{abstract}



\maketitle

\section{ Introduction}

Investigation of the properties of quark gluon plasma, a deconfined
 strongly interacting matter, remains a major activity of present day
high energy nuclear physics. This holds out a promise of a deeper
understanding of the laws of Quantum Chromo-Dynamics (QCD) and of
early universe which was in the form of quark-gluon plasma. Relativistic
heavy ion collisions studied at the Brookhaven National Laboratory
have produced extremely valuable results which confirm elliptic flow of
hadrons~\cite{flow}, suppression of hadrons having large transverse momenta or
jet-quenching~\cite{jet}, recombination of partons as a mechanism
of hadronization~\cite{recomb}, radiation of thermal photons~\cite{phot},
and suppression~\cite{helmut} and regeneration~\cite{bob} of
J/$\psi$ (see, Ref.~\cite{jpsi}). These findings clearly confirm formation of
quark gluon plasma, which is strongly interacting and has almost a vanishing
viscosity~\cite{roy}.

We are now ready to explore the next and more involved questions about the
quark-gluon plasma  in greater detail and precision. These questions,
in no particular order, could be the following.
When and how quickly
does the plasma thermalize? Does the time of thermalization depends on
the flavors of the partons?
What is the flavor dependence of energy loss of
partons? How does the plasma hadronize? How does it evolve in space and time?
How does the restoration of the chiral symmetry  upon formation of the
plasma and its breaking upon hadronization affect the evolution of the system?
 How does it break-up? What is the order of phase transition? Is there a
tri-critical point? Where does it lie? Shall we witness some of the more
exotic signatures of the formation of the plasma like the local CP
violation and chiral condensates?

A related set of questions concern the production and propagation of the
heavy quarks in such collisions. Heavy quarks offer some very distinct
advantages. They are mainly produced from prompt fusion of gluons
($gg \rightarrow Q\overline{Q}$) and quark-antiquark annihilation
($q\overline{q}\rightarrow Q\overline{Q}$). These processes can be accurately
described up to next to leading order using perturbative QCD. Their large
mass and the necessity to produce them as a $Q\overline{Q}$ pair, ensures
that their production at later times could be limited. Again, due to their very
large mass they move slowly, some-what like Gulliver in the Land
of Lilliput, through
the partonic wind of light quarks and gluons. They
will lose energy as they are buffeted by the partons; through collisions
and radiation of gluons. Several studies (see, e.g.~\cite{jamil} and references
therein) have made detailed evaluations for this energy loss.

Consider b-quarks moving through the plasma. Due to their
large mass it is expected that
while they will lose energy and momentum, the direction of their motion
may not change substantially due to these interactions which may involve
small momentum transfers, individually. Their momentum
does not undergo a large change as they fragment or coalesce with a light quark
to form B-mesons. The B-mesons may also not change their direction
drastically due to their interaction with pions. If true, this should make them
a valuable probe for the reaction plain dependence of the properties of the
plasma, especially for non-central collisions. Similar considerations may
apply to charm quarks as well, to a great extent. It has also been
 suggested that
the partonic flow may sensitively affect the back to back correlations of
heavy-quarks~\cite{nuxu1,nuxu2}.

The modification of the back-to-back correlations reported earlier
considers production at lowest order~\cite{nuxu1} and also at
NLO~\cite{nuxu2}, the later
contributing substantially to the production of charm quarks
at LHC. It is quite clear that
the heavy-quarks produced from splitting of gluons, for example would not move
back-to-back.

If other mechanisms, e.g., a thermal production of heavy-quarks,
or a pre-equilibrium
production of heavy-quarks due to interaction between partons having
large transverse
momenta, or a production due to passage of a parton having a
large transverse momentum (jet)
through the thermalized QGP makes a substantial contribution,
 then these back-to-back
correlations will be affected strongly. Of course the so called
nuclear modification factor
R$_{\rm {AA}}$ will have to be accounted for. It is also expected
 that the relative importance
of these contributions will depend on the transverse momentum
 distribution of the heavy quarks,
which will introduce further richness in these studies.
The correlated decay of charm and bottom mesons is also known to lead to a
substantial contribution to dileptons~\cite{ramona},
which have long been considered a reliable
signature of quark gluon plasma~\cite{shuryak}.

In order to put these possibilities to a rigorous test, as a first
step we consider
production of charm quarks at RHIC and LHC energies and of bottom quarks
 at LHC energies
due to prompt interactions, thermal productions, and pre-equilibrium productions
due to interaction of two partonic jets and due to passage of a partonic jet
through the quark gluon plasma. Our results suggests that there is a substantial
production of charm quarks at LHC following the initial (prompt) interaction.

The paper is organized as follows. In the next section we give
the formulations of
prompt, jet-jet, jet-thermal, and thermal interactions leading to
 production of heavy
quarks. We shall also consider a scenario, where the initially produced partons
undergo free-streaming, as an alternative to fully thermalize expansion.
The results are discussed in Sect. 4.
Finally in Sec. 5, we give our conclusions.
\begin{figure}[h]
\begin{center}
\includegraphics[scale=0.5]{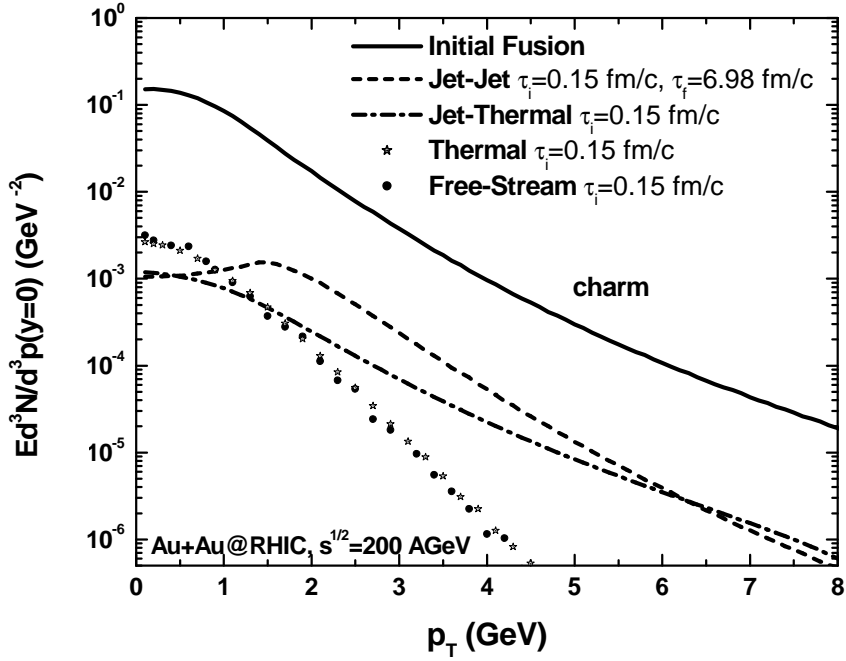}\\
\caption{The p$_{T}$ distribution for charm production from initial
 fusion (solid curve), jet-jet (dashed curve), jet-thermal (dash-dotted curve),
thermal (stars), and free-streaming (solid circles) processes
 with initial time 0.15 fm/$c$,
in central collision of gold nuclei at RHIC at $\sqrt{s}$ = 200 AGeV.}
\label{fig1}
\end{center}
\end{figure}

\begin{figure}[h]
\begin{center}
\includegraphics[scale=0.5]{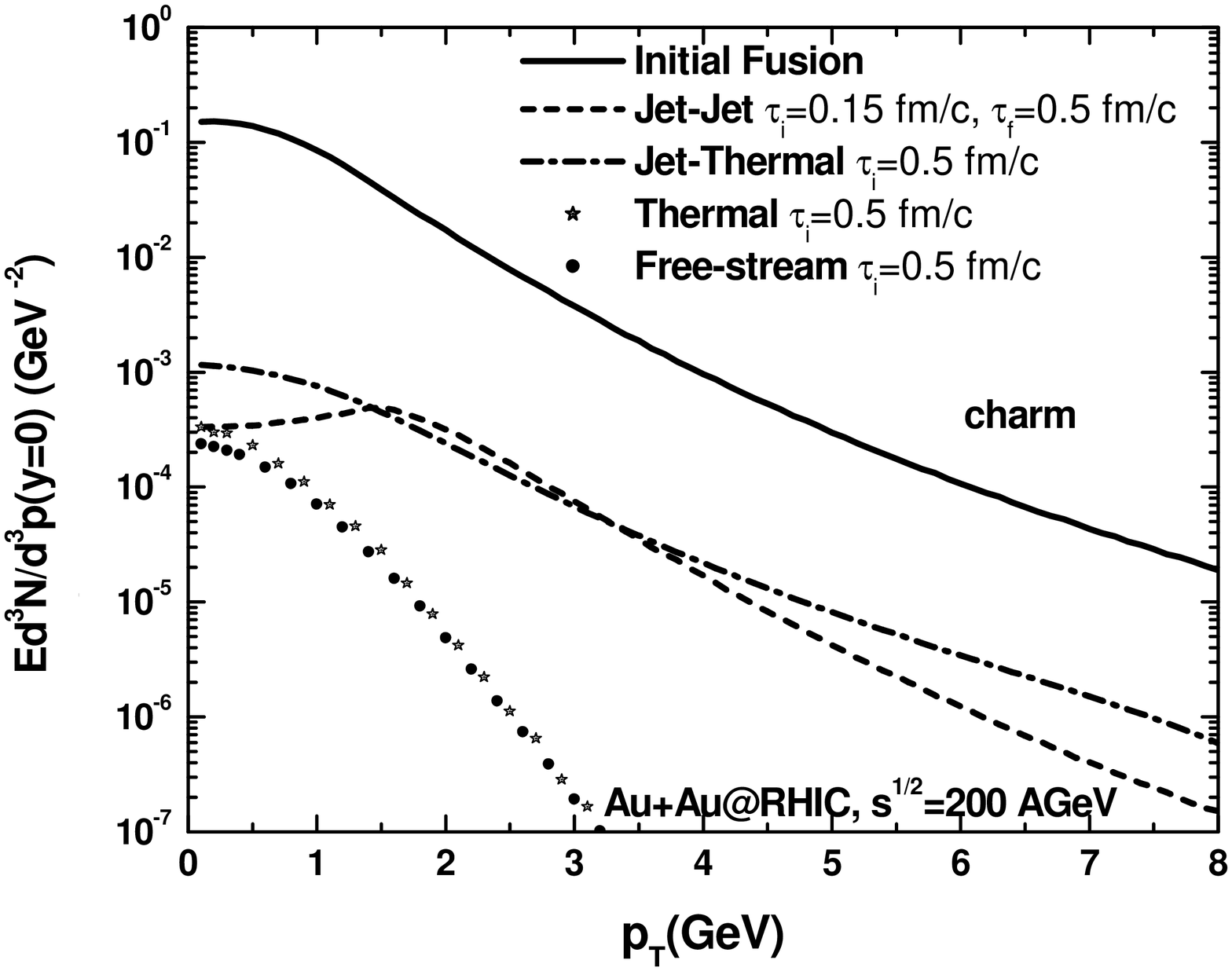}\\
\caption{The p$_{T}$ distribution for charm production from initial
 fusion (solid curve), jet-jet (dashed curve), jet-thermal (dash-dotted curve),
thermal (stars), and free-streaming (solid circles) processes
with initial time 0.5 fm/$c$,
in central collision of gold nuclei at RHIC at $\sqrt{s}$ =  200 AGeV.
The jet-jet contribution in this is thus truly pre-thermal.}
\label{fig2}
\end{center}
\end{figure}

\begin{figure}[h]
\begin{center}
\includegraphics[scale=0.5]{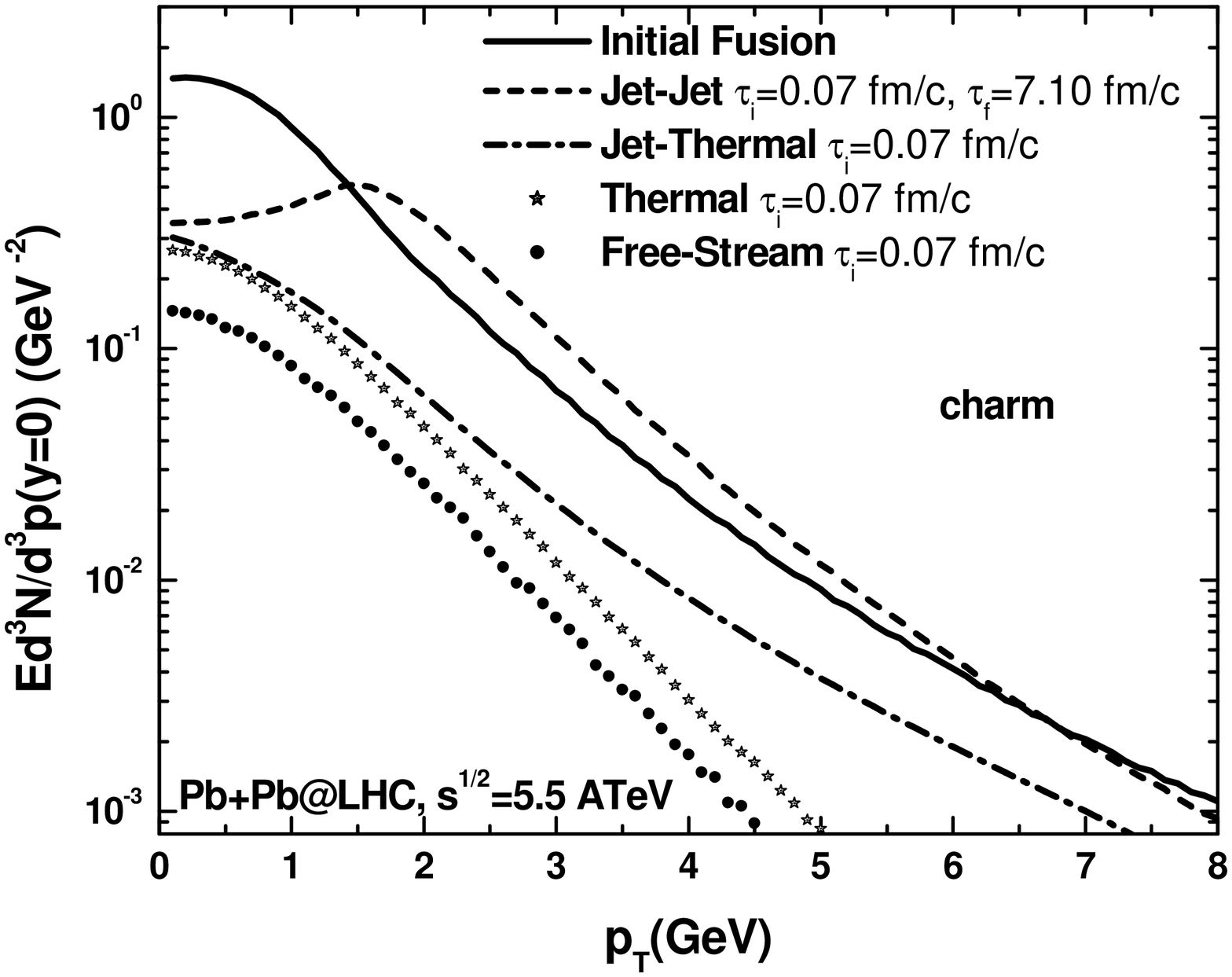}\\
\caption{The p$_{T}$ distribution for charm production from initial
 fusion (solid curve), jet-jet (dashed curve), jet-thermal (dash-dotted curve),
thermal (stars), and free-streaming (solid circles) processes
 with initial time 0.07 fm/$c$,
in central collision of lead nuclei at LHC at $\sqrt{s}$ =  5500 AGeV.}
\label{fig3}
\end{center}
\end{figure}

\begin{figure}[h]
\begin{center}
\includegraphics[scale=0.5]{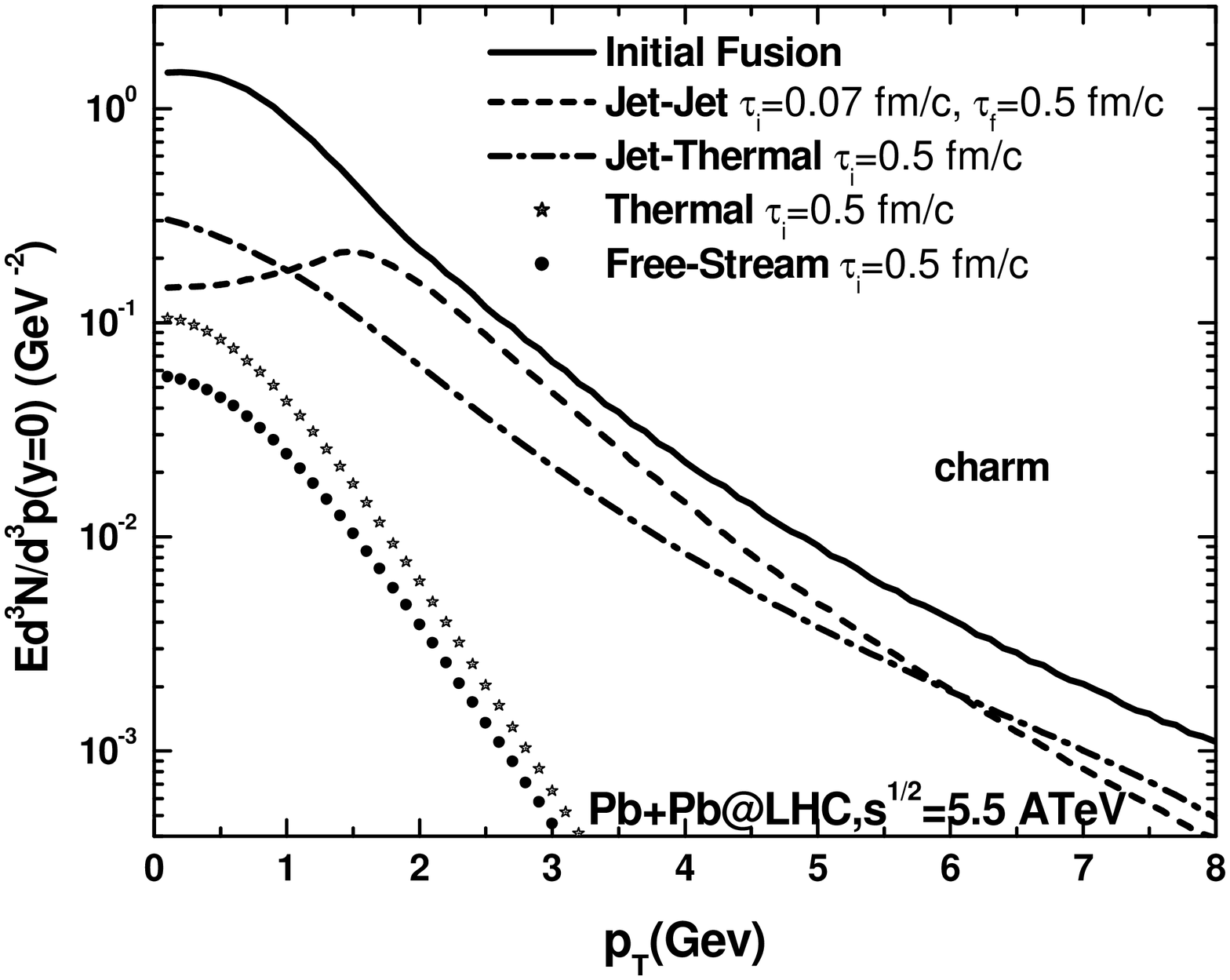}\\
\caption{The p$_{T}$ distribution for charm production from initial
 fusion (solid curve), jet-jet (dashed curve), jet-thermal (dash-dotted curve),
thermal (stars), and free-streaming (solid circles) processes
with initial time 0.5 fm/$c$,
in central collision of lead nuclei at LHC at $\sqrt{s}$ = 5500 AGeV.
The jet-jet contribution in this is thus truly pre-thermal.}
\label{fig4}
\end{center}
\end{figure}

\begin{figure}[h]
\begin{center}
\includegraphics[scale=0.47]{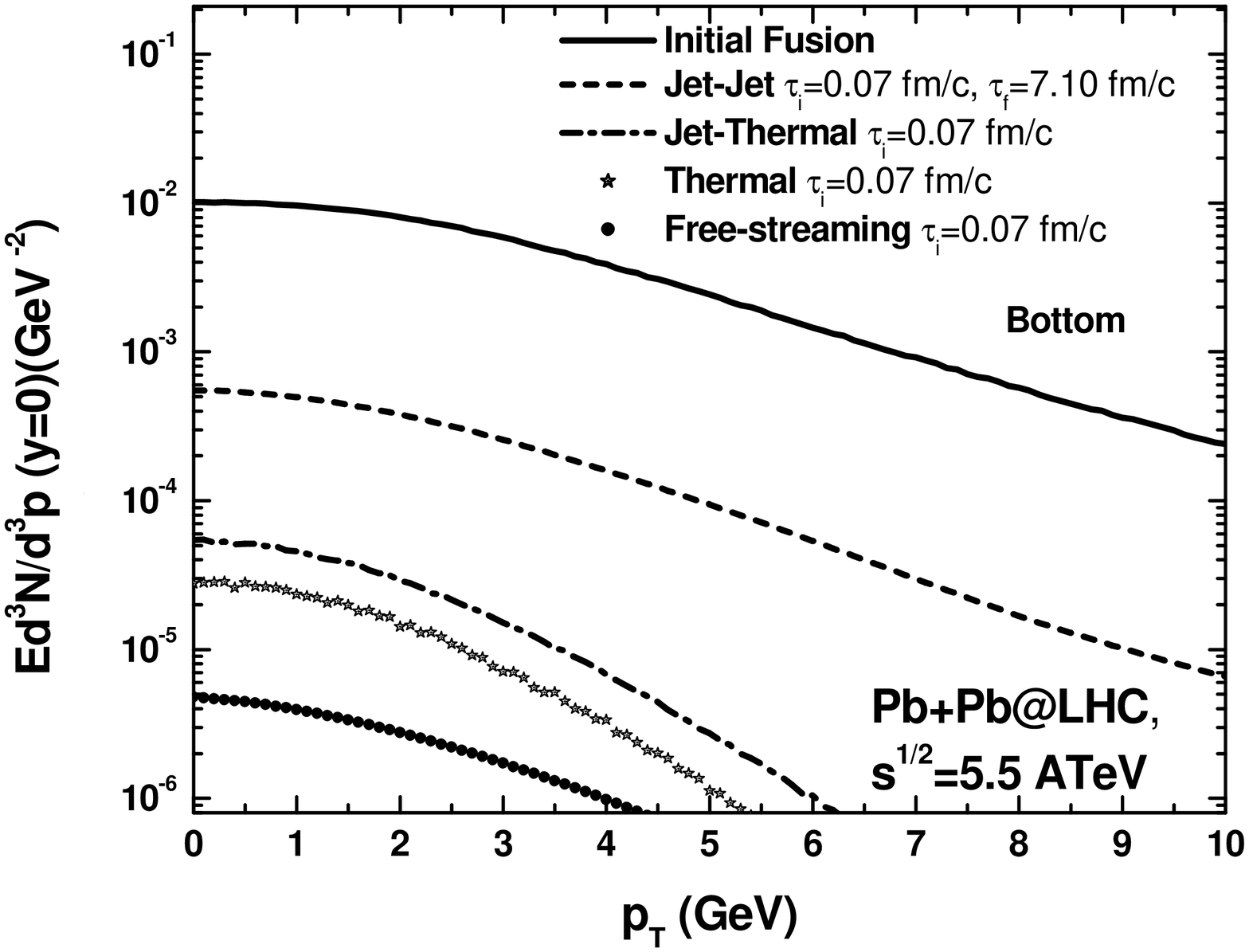}\\
\caption{The p$_{T}$ distribution for bottom production from initial
 fusion (solid curve), jet-jet (dashed curve), jet-thermal (dash-dotted curve),
thermal (stars), and free-streaming (solid circles) processes
with initial time 0.07 fm/$c$,
in central collision of lead nuclei at LHC at $\sqrt{s}$  =  5500 AGeV.}
\label{fig5}
\end{center}
\end{figure}

\begin{figure}[h]
\begin{center}
\includegraphics[scale=0.47]{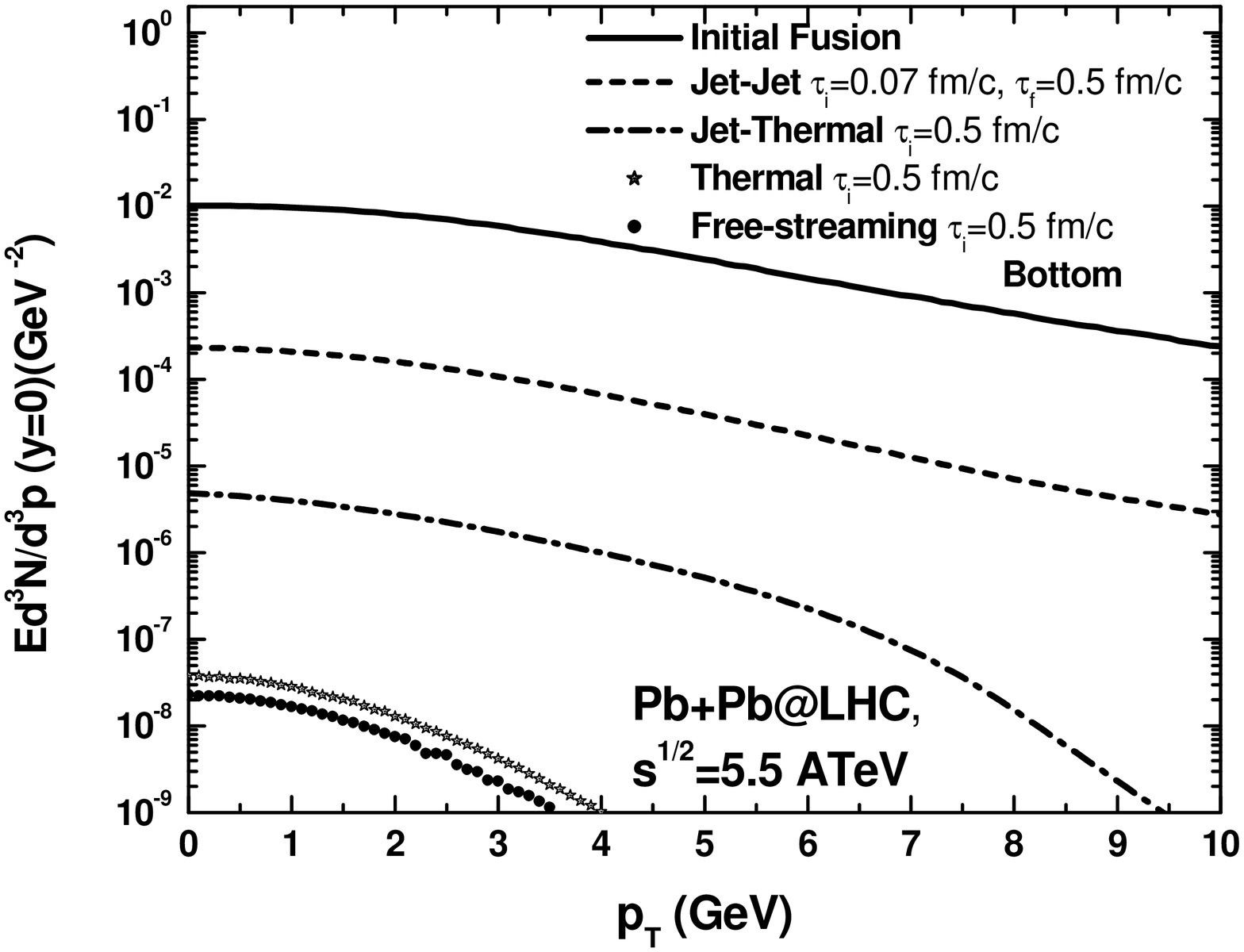}\\
\caption{The p$_{T}$ distribution for bottom production from initial
 fusion (solid curve), jet-jet (dashed curve), jet-thermal (dash-dotted curve),
thermal (stars), and free-streaming (solid circles) processes
 with initial time 0.5 fm/$c$,
in central collision of lead nuclei at LHC at $\sqrt{s}$ = 5500 AGeV.
The jet-jet contribution in this is thus truly pre-thermal.}
\label{fig6}
\end{center}
\end{figure}
\section{Initial Fusion (Prompt Interaction)}

Heavy ion collisions at RHIC (gold on gold) or at LHC (lead on lead) lead to
heavy quark productions primarily through gluon fusion
($gg \rightarrow Q \overline{Q}$) and also from light quark annihilation
($q \bar{q} \rightarrow Q \overline{Q}$).
The flavor excitations for intrinsic heavy quarks ($q Q \rightarrow q Q$ or
$g Q \rightarrow g Q$) is suppressed when next-to leading order
processes are considered (see, Ref.~\cite{collins}).

The differential cross-section for $gg \rightarrow Q \overline{Q}$ and
 $q \bar{q} \rightarrow Q \overline{Q}$ can be written as:

\begin{equation}
\frac{d\sigma}{d\hat{t}} = \frac{\left|M\right|^{2}}{16\pi\hat{s}^{2}}
\end{equation}
where the invariant amplitude $\left|M\right|^{2}$ is given by~\cite{combridge}:
\begin{equation}
\left|M\right|^{2}_{q \bar{q}\rightarrow Q \overline{Q}}=
 \frac{64\pi^{2}\alpha_{s}^{2}}{9}\left[\frac{(M^{2}-
\hat{t})^{2}+(M^{2}-\hat{u})^{2}+2M^{2}\hat{s}}{\hat{s}^{2}}\right]~,
\end{equation}
and
\begin{eqnarray}
\left|M\right|^{2}_{gg\rightarrow Q \overline{Q}} = \pi^{2}\alpha_{s}^{2}
& &\left[\frac{12}{\hat{s}^{2}}(M^{2}-\hat{t})(M^{2}-\hat{u})\right.\nonumber\\
&&+ \frac{8}{3}\frac{(M^{2}-\hat{t})(M^{2}-\hat{u})-2M^{2}(M^{2}+
\hat{t})}{(M^{2}-\hat{t})^{2}}\nonumber\\
&&+ \frac{8}{3}\frac{(M^{2}-\hat{t})(M^{2}-\hat{u})-2M^{2}(M^{2}+
\hat{u})}{(M^{2}-\hat{u})^{2}}\nonumber\\
&&- \frac{2}{3}\frac{M^{2}(\hat{s}-4M^{2})}{(M^{2}-\hat{t})(M^{2}-
\hat{u})}\nonumber\\
&&- 6\frac{(M^{2}-\hat{t})(M^{2}-\hat{u})+M^{2}(\hat{u}-\hat{t})}
{\hat{s}(M^{2}-\hat{t})}\nonumber\\
&&- \left. 6\frac{(M^{2}-\hat{t})(M^{2}-\hat{u})+M^{2}
(\hat{t}-\hat{u})}
{\hat{s}(M^{2}-\hat{u})}\right]~.
\end{eqnarray}

The running coupling constant $\alpha_{s}$ for lowest order is given by

\begin{equation}
\alpha_{s}=\frac{12\pi}{(33-2N_{f})\ln(Q^{2}/\Lambda^{2})}.
\end{equation}
where $N_{f}$=3 is the number of flavors and $\Lambda$ is the QCD scale
parameter used in the running coupling constant.

The cross-section for the production of heavy quarks from proton-proton
collisions at leading order~\cite{ehlq} is given by

\begin{eqnarray}
\frac{d\sigma}{dy_{1} dy_{2} d^{2}p_{T}} &=& 2x_{1}x_{2}\sum_{ij}
\left[f^{(1)}_{i}(x_{1},Q^{2})f_{j}^{(2)}(x_{2},Q^{2})
\frac{d\hat{\sigma}_{ij}(\hat{s},\hat{t},\hat{u})}{d\hat{t}}
\right.\nonumber\\
&+& \left.f_{j}^{(1)}(x_{1},Q^{2})f_{i}^{(2)}(x_{2},Q^{2})
\frac{d\hat{\sigma}_{ji}(\hat{s},\hat{u},\hat{t})}{d\hat{t}}\right]
/(1+\delta_{ij})~,
\end{eqnarray}
where i and j are the interacting partons and
$f_{i}$ and $f_{j}$ are the partonic structure functions, and
$x_{1}$ and $x_{2}$ are the momentum fractions of the
parent nucleons carried by the
 interacting partons.
 The effect of nuclear shadowing which becomes more
pronounced at small $x$ has been included using the EKS98~\cite{eks98}
parametrization. We use
 CTEQ5L structure function set for nucleons~\cite{cteq5l}.
 The intrinsic transverse momentum of the interacting
 partons is neglected.

The light quark (u, d, s) and gluon production is similarly
 calculated with $d\sigma/d\hat{t}$ taken from Ref.~\cite{owens} with
 vanishing quark masses.

 For heavy ion collisions the $p_{T}$
spectrum for heavy quark production is given by

\begin{equation}
\frac{dN}{d^{2}p_{T}dy}= T_{AA}\frac{d\sigma}{d^{2}p_{T}dy}
\end{equation}
where for central collisions,
 $T_{AA}$=286 fm$^{-2}$ for Au+Au at RHIC and $T_{AA}$=292 fm$^{-2}$
for Pb+Pb at LHC. We  account for higher order corrections
 by taking a constant K-factor $\approx$ 2.5 (see, Ref.~\cite{jamil}).

\section{Interaction among Partons}

The initial hard scattering between the partons of the two nuclei
 will lead to gluons and light quarks having large transverse momenta.
 These will ultimately fragment and
 lead to a stream of hadrons into a narrow cone
 or jets. We shall continue to call gluons and light quarks having
 large transverse momenta as jet particles. Being copious in number,
 they would interact again and may even approach thermalization.
The first collisions among two gluonic jets or a quark and
anti-quark jet are likely to have sufficient energy to produce
a pair of heavy quarks. Thus it is felt that jet-jet interaction
 in relativistic heavy ion collision can be an interesting source
 of production of heavy quarks~\cite{lg95}.

Several authors have predicted that very large initial temperatures
could be attained at RHIC~\cite{phot} and LHC energies~\cite{kms92}.
 This suggests that even
 a thermal production of heavy quarks through the interaction of
 thermalized quarks and gluons can take place.

An interesting new source of high momentum photons~\cite{fms03}
 and dileptons~\cite{fgs03} has
 recently been proposed, which arises from the passage of high
 momentum quark or gluon jets through the thermalized quark
gluon plasma. It is of interest to see it whether the same
 process could lead to production of heavy quarks.
 We shall call this process as jet conversion or jet-thermal
interaction~\cite{lf08}.

The general expression for the production of a heavy quark at
central rapidity is given by~\cite{lg95,lmw}:

\begin{eqnarray}
\left.E\frac{d^{3}N}{d^{3}p}\right|_{y=0} &=& \int d^{4}x
\int\frac{1}{16(2\pi)^{8}}\frac{d^{3}p_{1}d^{3}p_{2}d^{3}p^{'}}
{\omega_{1}\omega_{2}}\nonumber\\
&\times&\delta^{4}\left(\sum p^{\mu}\right)/E'\nonumber\\
&\times&\left|M\right|^{2} F(\vec{x},\vec{p_{1}},t)F(\vec{x_{2}},
\vec{p_{2}},t)\nonumber\\
\end{eqnarray}
where $\sum p^{\mu}= p_{1}+p_{2}-p-p'$,  $p_{1}$ and
 $p_{2}$ are the four-momenta of the incoming partons and
 $p$ and $p'$ are the same for outgoing heavy quark and
anti-quark. $F(\vec{x}, \vec{p}, t)$ gives the
phase space distribution function for the incoming partons.

Before deriving results for different processes by choosing
 appropriate phase space distributions etc., we can perform
 the following simplifications:

Writing $d^{3}p_{i}/\omega_{i}=p_{T_{i}}dp_{T_{i}}d\phi_{i}dy_{i}$
and integrating over $d^{3}p'$, we get for $y$=0 ($p_{z}$=0,
and $p_{T}$=$p$):
\begin{eqnarray}
\int\frac{d^{3}p_{1}d^{3}p_{2}d^{3}p'}{\omega_{1}\omega_{2}E'}
\delta^{4}\left(\sum p^{\mu}\right)\left|M\right|^{2}
 F(\vec{x},\vec{p_{1}},t)F(\vec{x},\vec{p_{2}},t)\nonumber\\
=\int dy_{1}~dy_{2}~d\phi_{1}~d\phi_{2}~p_{T2}dp_{T2}~p_{T1}dp_{T1}
\frac{\delta(\sum E)}{E'}\left|M\right|^{2}\nonumber\\
\times F(\vec{x},p_{T1},\phi_{1},y_{1},t)F(\vec{x},p_{T2},
\phi_{2},y_{2},t)~,\nonumber\\
\label{master}
\end{eqnarray}
where
\begin{equation}
\frac{\delta(\sum E)}{E'}= \frac{\delta (p_{T1}-p_{T1,0})}{
\left[p_{T2}(\cosh(y_{1}-y_{2})\right.
-\left.\cos(\phi_{1}-\phi_{2}))-(E\cosh y_{1}-p\cos\phi_{1})
\right]}~,\nonumber\\
\label{debyep}
\end{equation}
and we have
\begin{equation}
p_{T1,0}=\frac{p_{T2}(E\cosh y_{2}-p\cos\phi_{2})}{[p_{T2}(\cosh (y_{1}-y_{2})
-\cos (\phi_{1}-\phi_{2}))-(E\cosh y_{1}-p\cos\phi_{1})]}~.\nonumber\\
\label{pt10}
\end{equation}

Now we proceed to evaluate individual contributions.

\subsection{Jet-Jet interaction of partons}

In order to estimate the contribution of the jet-jet interaction to the
production of heavy quarks, we approximate~\cite{lg95} the
phase-space distribution of the gluon, quark, or anti-quark jets
produced in initial
(prompt) scattering of the partons in a central collision as:

\begin{equation}
F(\vec{x},\vec{p},t)= f_{\rm {jet}}(\vec{x},\vec{p},t)~,
\label{jet1}
\end{equation}
where ~\cite{lg95}
\begin{equation}
f^{i}_{\rm {jet}}(\vec{x},\vec{p},t) = \frac{(2\pi)^{3}}{g_{i}
\tau\pi R_{T}^{2}p_{T}} \frac{dN_{i}}{dyd^{2}p_{T}}
\, \delta(y-\eta)\,\Theta(\tau_{f}-\tau)
\, \Theta(\tau-\tau_{i})~,
\label{jet2}
\end{equation}
with $p_{T} >$ 2 GeV.
Here $\delta(y-\eta)$ denotes the Bjorken correlation for space-time
and energy-momentum rapidities and '$i$' stands for quarks, anti-quarks,
 or gluons. The degeneracy of quarks and gluons is given by $g_{g/q}$
 such that $g_{g}=8\times2$ and $g_{q} =3\times2$~\cite{rm}. Further
$R_{T}$ is the transverse radius of the nucleus and $dN_{i}/d^{2}p_{T}dy$
 is the transverse momentum distribution of partons for $p_{T} >$ 2 GeV.
We neglect the dependence of this distribution on the momentum
rapidity~\cite{lg95} as we are calculating the results for heavy quarks at y=0,
when only very small values of y$_{1}$ and y$_{2}$ contribute, and the
rapidity dependence is marginal.
 For results at y~$\neq$~0, appropriate distributions will need to be used.

The momentum space distribution of the jets at RHIC and LHC are taken from
parametrization~\cite{fgs03} given earlier, where the jet distributions
 were calculated in LO-pQCD with a K-factor ($\approx$ 2.5), CTEQ5L
 structure functions and EKS98 shadowing functions, which we have
 used here for calculation of initial production of heavy quarks.
Thus we have,
\begin{equation}
\frac{dN}{dyd^{2}p_{T}} = \left.T_{\rm{AA}}
\frac{d\sigma^{\rm {jet}}}{d^{2}p_{T}dy}\right|_{y=0}
                    =K\,\frac{C}{(1+p_{T}/B)^{\beta}}~,
\end{equation}
and
\begin{equation}
h^{i}_{\rm {jet}}(p_{T})=\frac{1}{g_{i}}\left.
\frac{dN}{d^{2}p_{T}dy}\right|_{y=0}~,
\end{equation}
where K, C, B, and $\beta$ are taken from reference~\cite{fgs03}.

We neglect transverse expansion of the system, which should be
valid at early times when most of the heavy quarks are produced.
Now taking
\begin{equation}
d^{4}x= \tau\,d\tau\,r\,dr\,d\eta\,d\phi_{r}~,
\end{equation}
we can perform the integration over $r$, $\phi_{r}$ and $\tau$.

Thus the $p_{T}$ distribution of open heavy quark production
from jet-jet interaction can be written as
\begin{eqnarray}
\left.E\frac{d^{3}N}{d^{3}p}\right|_{y=0}&=& \frac{\ln(\tau_{f}/
\tau_{i})}{16(2\pi)^{2}(\pi R_{T}^{2})}\int
 d\eta ~dp_{T2}~d\phi_{1}~d\phi_{2}\nonumber\\
& & \times\left[\frac{1}{p_{T2}(1-\cos(\phi_{1}-\phi_{2}))-
(E\cosh\eta-p\cos\phi_{1})}\right]\nonumber\\
& &\times\left[\frac{1}{2}g^{2}_{g}h^{g}_{\rm {jet}}(p_{T1,0})
h^{g}_{{\rm {jet}}}
(p_{T2})
\left|M\right|^{2}_{gg\rightarrow Q\overline{Q}}\right.\nonumber\\
&&\,\,\,\,\,\,\,
\left. + g^{2}_{q}N_{f}h^{q}_{\rm {jet}}(p_{T1,0})h^{q}_{\rm {jet}}(p_{T2})
\left|M\right|^{2}_{q\bar{q}\rightarrow Q\overline{Q}}\right]~,
\label{jet-jet}
\end{eqnarray}
where $N_{f}$ = 3 is the number of quark flavors~\cite{comment}.

While writing the above, we have used 
the Bjorken correlations $\delta(y_{1}-\eta)$ and
 $\delta(y_{2}-\eta)$ (see Eq.\ref{jet2}). With this, the 
 Eqs.(\ref{debyep}) and (\ref{pt10})
reduce to,
\begin{equation}
\frac{\delta(\sum E)}{E'}=\frac{\delta(p_{T1}-p_{T1,0})}{p_{T2}(1-
\cos(\phi_{1}-\phi_{2}))-(E\cosh\eta-p\cos\phi_{1})}~,
\end{equation}
and
\begin{equation}
p_{T1,0}=\frac{p_{T2}(E\cosh\eta-p\cos\phi_{2})}{p_{T2}(1-
\cos(\phi_{1}-\phi_{2}))-(E\cosh\eta-p\cos\phi_{1})}~,
\end{equation}
and thus Eq.(\ref{master}) reduces to Eq.(\ref{jet-jet}), above.
Numerical integration of the Eq.\ref{jet-jet} gives pre-thermal
 heavy quark production. We shall see that the major contribution to heavy
quark production comes from gluon fusion.

We add that Levai {\em et al.}~\cite{lmw}
calculated the so-called pre-equilibrium (jet-jet) contribution
by assuming that this mechanism operates during
$\tau$ $\epsilon $~[0.1 -- 0.5]~fm/$c$.
We shall explore some other options as well.

We re-iterate that the above expressions do not account for energy
loss of the energetic gluons  and quarks as they traverse the plasma
and thus they provide the upper limit of the heavy quark productions.

\subsection{Interaction among thermally and chemically
equilibrated partons }

If multiple scatterings occur in a rapid succession, QGP may reach
 thermal equilibrium quite early and follow hydrodynamics evolution
 till it hadronizes. During this phase of evolution, heavy quarks may
 be produced provided the initial temperature is high~\cite{shor}.

We can estimate the initial temperature by assuming Bjorken
hydrodynamics~\cite{bj} which relates it to the particle rapidity density by
\begin{equation}
\frac{2\pi^{4}}{45\zeta(3)\pi R_{T}^{2}} \frac{dN}{dy}=4aT_{0}^{3}\tau_{0}
\label{dnt}
\end{equation}
where $dN/dy$ is the particle rapidity density, $a$=$42.25\pi^{2}/90$
for massless light quarks and gluons and $R_{T}=1.2A^{1/3}$.
 We take particle rapidity density as 1260 estimated experimentally~\cite{phe}
 at RHIC and assumed~\cite{kms92} that the
 particle rapidity density at LHC is about 5625. Some recent works
suggest a smaller value for $dN/dy \approx$  3000--3500
 at LHC~\cite{dima} from considerations
of parton saturation.  However, larger values have also
been suggested~\cite{nestor}.
 The initial experimental
 results from $pp$ collisions at
LHC at 0.9 TeV,  2.36, and 7 TeV already show an increase
in the particle rapidity densities, which is steeper than
expected~\cite{cms}. In any case, results for any other rapidity
density can be easily obtained.

The time evolution of the temperature of thermalized QGP
for a boost-invariant longitudinal expansion is governed by~\cite{bj}:
\begin{equation}
T_{0}^{3}\tau_{0}=T^{3}\tau= \rm{const}.
\end{equation}
Assuming a rapid thermalization and chemical equilibration,
we get the lowest estimate of $\tau_{0}$ from the above by
 taking $\tau_{0}\approx$ 1/3T$_{0}$ ~\cite{kms92}.
 This provides that $\tau_{0}\approx$ 0.15 fm/$c$ at RHIC.
Such a small value for $\tau_0$ is supported by the 
single photon data~\cite{phot}. 
As an alternative, we also consider a much larger time of
thermalization, $\tau_{0}\approx$ 0.5 fm/$c$, with $T_{0}$
calculated from Eq.~\ref{dnt}. We add that several studies, especially
the ones related to the flow of hadrons tend to use
 a larger value of $\tau_0\approx$ 0.6 fm/$c$~\cite{flow}, even
though more recent studies find only a weak dependence of $\tau_0$
on the flow for hadrons~\cite{rupa}.
For LHC, we have similarly assumed
 $\tau_{0}\approx$ 0.07 fm/$c$ and 0.5 fm/$c$.

Taking the critical temperature
 is taken to be 0.170 GeV,  we see that the end of the QGP phase occurs at:
\begin{equation}
\tau_{f}=({\rm {const.}}/0.170^{3})
\end{equation}
and then the thermal production mechanism would operate during
$\tau_0$ to $\tau_f$.

We take the phase space distribution for the thermalized quarks and gluons as,
\begin{equation}
f^{i}_{th}(p_{T},y,\eta)=\exp\left[-p_{T}\cosh(y-\eta)/T\right]~.
\end{equation}

Thus the transverse momentum distribution of thermally produced charm given by
\begin{eqnarray}
\left.E\frac{d^{3}N}{d^{3}p}\right|_{y=0} =
\frac{\pi R_{T}^{2}}{16(2\pi)^{8}}\int \, \tau \, d\tau \,
 d\eta \, dp_{T2} \, d\phi_{1} \,d\phi_{2} \, dy_{1} \, dy_{2}
\nonumber\\
\times\frac{(p_{T2}~p_{T1,0})}{[p_{T2}(\cosh(y_{1}-y_{2})-
\cos(\phi_{1}-\phi_{2}))
-(E\cosh y_{1}-p\cos\phi_{1})]}\nonumber\\
\times\left[f_{th}(p_{T1,0},y_{1},\eta)f_{th}(p_{T2},y_{2},\eta)\right]
\times\left[\frac{1}{2}g_{g}^{2}
\left|M_{gg\rightarrow \overline{Q}Q}\right|^{2}
+ g_{q}^{2}N_{_{f}}\left|M_{q\bar{q}\rightarrow
\overline{Q}Q}\right|^{2}\right]\nonumber\\
\label{thermal}
\end{eqnarray}

Just as in pre-thermal case, we
calculate the final charm production at y=0 or central rapidity and thus
 $p_{z}=0$ and $p_{T}=p$. We now have the kinematical constraint;

\begin{equation}
\frac{\delta(\sum E)}{E'} =\frac{\delta (p_{T1}-p_{T1,0})}{[p_{T2}(\cosh(y_{1}-y_{2})
-\cos(\phi_{1}-\phi_{2}))-(E\cosh y_{1}-p\cos\phi_{1})]}\nonumber\\
\end{equation}
and
\begin{equation}
p_{T1,0}=\frac{(p_{T2}(E\cosh y_{2}-p\cos\phi_{2}))}{[p_{T2}(\cosh (y_{1}-y_{2})
-\cos (\phi_{1}-\phi_{2}))-(E\cosh y_{1}-p\cos\phi_{1})]}\nonumber\\
\end{equation}

Numerical integration of the Eq.~\ref{thermal} for different initial conditions
gives us the contribution from thermalized QGP.

\subsection{Interactions of free-streaming partons}

As an extreme, we consider free-streaming partons, as a model of
 evolution of the system of deconfined quarks and gluons, which
completely relaxes the condition of thermalization. The initial
 distribution at $t=\tau_{0}$ and $z=0$
is obtained by assuming maximum entropy ~\cite{kms92}, so that
\begin{equation}
f(p,x)=\frac{dN}{d^{3}pd^{3}x}=\exp(-\frac{E}{T_{o}})~,
\end{equation}
and the condition that needs to be satisfied is
\begin{equation}
p^{\mu}\frac{\partial f(x,p)}{\partial x^{\mu}}=0~.
\label{defer}
\end{equation}

We assume boost invariance along the z-axis with
\begin{equation}
f(p,x)=f(p_{T},p_{z}t-Ez)~.
\end{equation}

The solution which satisfies the differential Eq.~\ref{defer} is

\begin{equation}
f(p,x)=\exp\left[-\frac{\sqrt{p_{T}^{2}+(p_{z}t-Ez)^{2}/
\tau_{0}^{2}}}{T_{0}}\right]~.
\label{fpx}
\end{equation}

Now using
\begin{eqnarray}
p_{z}&=&p_{t}\sinh y,~ E=p_{t}\cosh y,\nonumber\\
z&=&\tau\sinh\eta,~ t=\tau\cosh\eta~,
\end{eqnarray}
Eq.~\ref{fpx} becomes
\begin{equation}
f(p_{T},\eta,y)=\exp\left[-\frac{p_{T}\sqrt{1+\tau^{2}\sinh^{2}(y-\eta)/
\tau_{0}^{2}}}{T_{0}}\right]~.
\end{equation}

Thus the final integration to calculate $p_{T}$ distribution for
heavy quark production  from free streaming partons is given by

\begin{eqnarray}
\left.E\frac{d^{3}N}{d^{3}p}\right|_{y=0} =\frac{\pi
 R_{T}^{2}}{16(2\pi)^{8}}\int\tau d\tau~
d\eta ~dp_{T2}~d\phi_{1}~d\phi_{2}~dy_{1}~dy_{2}\nonumber\\
\times\frac{(p_{T2}~p_{T1,0})}{[p_{T2}(\cosh(y_{1}-y_{2})-
\cos(\phi_{1}-\phi_{2}))-(E\cosh y_{1}-p\cos\phi_{1})]}\nonumber\\
\times\left[f(p_{T1,0},\eta_{1},y_{1})~f(p_{T2},\eta_{2},y_{2})\right]
\times\left[\frac{1}{2}g_{g}^{2}\left|M_{gg\rightarrow
\overline{Q}Q}\right|^{2}+ g_{q}^{2}N_{f}
\left|M_{q\bar{q}\rightarrow \overline{Q}Q}\right|^{2}\right]\nonumber\\
\end{eqnarray}
The initial conditions for the free-streaming case are taken to be
 same as that for the thermal production, whereas the final time is
taken as $R_{T}/c$, the transverse radius of the nuclei, after which the
system would surely expand rapidly along the transverse direction as
well and disintegrate.

\subsection{Charm production from the passage of jets through
thermal medium }

Now we discuss the production of heavy quarks by passage of light quark
 and gluonic jets through thermalized QGP. In order to proceed,
 we can differentiate the phase space distribution into jet partons
 and thermalized partons. The phase space distribution for equilibrated
 medium is given by

\begin{equation}
f_{th}=\exp\left[-p_{T}\cosh(y-\eta)/T\right]~.
\end{equation}

The jet distribution is given by Eq.~\ref{jet1} and \ref{jet2}
for $p_{T} >$ 2 GeV.
 We have already discussed $\tau_i$ or $\tau_0$, which gives the start
of the time from when we consider the system to be in the form of QGP.
We define $\tau_{d}$ as the time which a jet takes to reach the surface of
 the quark gluon plasma. Consider a jet formed  at $\vec{r}$ with velocity
$\vec{v}$ which travels
 to the surface of plasma. The distance, $d$, covered in this process
is given by,
\begin{equation}
d= -r\cos\phi+\sqrt{R_{T}^{2}-r^{2}\sin^{2}\phi}~,\\
\end{equation}
where $\phi=\cos^{-1}(\hat{v}\cdot\hat{r})$,
 and $R_{T}$ is the radius of the system.
A massless quark or a gluon would take a time
\begin{equation}
\tau_{d}= d/c
\end{equation}
for this journey.
Considering that QGP would cool down to the critical temperature by
 $\tau$ = $\tau_{f}$, the time spent by the jet in the plasma would be
 $\tau_{\rm {max}}$ = {\rm {min} $\left[\tau_{f},\tau_{d}\right]$ (see
Ref.~\cite{fgs03}),
 and the jet-thermal production mechanism would be in operation during
 $\tau$ $\epsilon$ [$\tau_{i}$, $\tau_{\rm {max}}$].

The thermalization time is taken as either 0.15 fm/$c$ or 0.5 fm/$c$
at RHIC and
as either 0.07 fm/$c$ or 0.5 fm/$c$ at LHC as before.

 The final result for
jet-thermal interaction is given by

\begin{eqnarray}
\left.E\frac{d^{3}N}{d^{3}p}\right|_{y=0}=\frac{1}{16(2\pi)^{4}\pi R_{T}^{2}}
\int d\tau~rdr~d\eta ~d\phi_{1}~d\phi_{2}~dy_{1}~dp_{T2}\nonumber\\
\times\frac{p_{T1,0}}{(p_{T2}(\cosh(y_{1}-\eta)-\cos(\phi_{1}-\phi_{2}))
-E\cosh y_{1}+p_{z}\sinh y_{1}+p_{T}\cos\phi_{1})}\nonumber\\
\times f_{th}(p_{T1,0},y_{1},\eta)
\left[g_{q}^{2}N_{f}h^{q}_{{\rm jet}}(p_{T2})\left|M\right|^{2}_{q\bar{q}\rightarrow Q\overline{Q}}
+\frac{1}{2}g_{g}^{2}h^{g}_{{\rm {jet}}}(p_{T2})\left|M\right|^{2}_{gg
\rightarrow Q\overline{Q}}\right]~,\nonumber\\
\label{free}
\end{eqnarray}
which we evaluate numerically.

We add that these results do not account for the energy loss suffered by the
high energy quark of gluon as it traverses the plasma, and during which it
may even change its flavour. This would necessitate a treatment along the
lines of Ref.~\cite{guy} for production of photons from a similar process.

\section{Results}

Now we discuss our results. As a first step
we show the results for charm production at RHIC and LHC.

\subsection{Charm at RHIC}

In Fig.~\ref{fig1}, we plot the results at RHIC for charm production from
prompt interactions (initial fusion), jet-jet interactions of partons, thermal
production and from the passage of high momentum jets through thermalized
quark-gluon plasma. We also give the free-streaming results to
compare with thermal production. We consider central collision
 of gold nuclei at $\sqrt{s}$ = 200 AGeV.

We see that the contribution of the initial fusion dominates at all
 $p_{T}$. Considering initial formation time $\tau_{0}$ as 0.15 fm/$c$,
 we can consider two extremes for the jet-jet interactions as
indicated earlier. For one extreme we consider that jet-jet
interaction continues till jets reach the surface of the system
 ($\tau_{f}$ $\approx$ $R_{T}/c$). The other extreme could be
 to assume that it operates only till the time of thermalization,
 if it is large ($\tau_{f}$ $\approx$ 0.5 fm/$c$), so that it is
considered as the pre-equilibrium contribution~\cite{lmw}. In any case the
 time integration for this contribution reduces to
$\ln(\tau_{f}/\tau_{i})$,(see Eq.~\ref{jet-jet}) and thus one can easily
 obtain this results for any choice of initial conditions.

Our results for this contribution at RHIC are similar in magnitude and
form to those reported by Lin and Gyulassy~\cite{lg95} for RHIC energies.
Recall that we consider only partons having $p_{T} >$ 2 GeV as
constituting the jets. The jet-thermal contribution is seen to
 be comparable to the contribution of the jet-jet interaction.

We find that the contribution of thermal production at large $p_{T}$
is rather small. However it is larger by a factor of about 3 at
lower $p_{T}$, compared to the jet-jet and jet-thermal contributions.
 This has  its origin in large initial temperature for low
$\tau_{0}$ value assumed here and exclusion of partons having
$p_{T} <$ 2 GeV in the jet distribution. The contribution obtained
by assuming free-streaming partons having same initial  $\tau_{0}$ and
operating till $\tau_{f} = R_{T}/c$ is quite similar to thermal
contribution. We also note that both thermal and free-streaming
contributions drop a lot more rapidly for larger $p_{T}$ compared
 to contributions initiated by prompt and jet interactions.
The latter contribute at a level of about 4-5 $\%$ for large
 $p_{T}$ as compared to the initial fusion for $\tau_{0} =$ 0.15 fm/$c$
and $T_{0} = $ 447 MeV. We add that a much larger initial temperature of
 700 MeV and K=4  has been assumed
in the calculations by Liu and Fries~\cite{lf08}
 for the jet conversion in thermalized medium.

In Fig.\ref{fig2}, we give results obtained by using the formation time
 of the plasma as 0.5 fm/$c$, and further assuming that the jet-jet
 interaction is in operation during $\tau \approx$ 0.15 fm/$c$ to 0.5 fm/$c$,
 so that it can be considered as a pre-equilibrium contribution~\cite{lmw}.
 In any case, we find that initial fusion gives the largest
 contribution to charm production, followed by jet-jet and
jet-thermal interactions. At high $p_{T}$, jet-jet as well as
jet-thermal interactions give  a contribution of about 1$\%$
of the initial fusion.
Now the thermal and free-streaming contribution though remaining
similar in magnitude, are much smaller and also fall more rapidly
with $p_{T}$.  They give a contribution at a level of
1$\%$ of the initial fusion contribution at $p_{T}$ $<$ 3 GeV.

We realize that even though small in magnitude, the jet-jet and
jet-thermal contributions may still give a discernible feature to
 the back-to-back correlation of charm-quarks at RHIC.
The correlation of charm quarks from these should be distinct
 from initial fusion (back-to-back for LO contribution and
forward peaked for NLO contribution). This aspect is under study.

We have not given our results for bottom production at RHIC energies here
as we have found that bottom production from the jet-jet mechanism
is less than 3 orders of magnitude, the thermal mechanism is less than
6 orders of magnitude and the jet-thermal mechanism is less than 4 orders
of magnitude of that obtained from initial (prompt) interaction~\cite{jamil}.
We do add, however, that the decay of bottom into charm quarks adds an 
interesting richness to the study of charm production as well 
as charm-correlation.

\subsection{Charm and bottom at LHC}

Next we discuss our results for charm and bottom production at LHC.

We consider central collision of lead nuclei at $\sqrt{s}$ = 5500 AGeV.
 Other initial conditions have already been discussed.
We find that (see Fig.\ref{fig3}) the charm production from initial
fusion is about a factor of 10 or more than that at RHIC, and
of-course its fall with $p_{T}$ is considerably slower, as one would expect.

We plot our results at LHC, taking time of formation of QGP to be
 0.07 fm/$c$. We have assumed the initial time of jet-jet interaction
to be from $\tau_{i}< $ 0.1 fm/$c$ until the jets reach the surface of
 the system, $\tau_{f} \approx R_{T}/c$ as one of the extremes.
We find jet-jet contribution even exceeds the initial fusion for
 $p_{T} >$ 2 GeV. Of-course since we include jets having $p_{T}>$ 2 GeV,
this contribution decreases for lower $p_T$ . We shall return to this.
The thermal production of the charm, taking advantage of the initial
 temperature when $\tau_{i} \approx$ 0.07 fm/$c$, is almost 40$\%$ of
the initial fusion contribution at low $p_{T}$, and drops to about
 5--10$\%$ of the same at $p_{T}$ $\approx$ 5 GeV. The jet-thermal
contribution is also seen to be considerable at large $p_{T}$, though
smaller than jet-jet contribution. It is seen to be comparable to the thermal
 contribution at lower $p_{T}$. Of-course this has its origin  in the large
initial temperature.

Next in Fig.\ref{fig4}, we plot our results at LHC for another set of initial
conditions. Taking time of equilibration to be 0.5 fm/$c$, we have
 pre-thermal charm production resulting from jet-jet
 interactions starting  at $\tau_{i}=$ 0.07 fm/$c$
and operating till  the time of thermalization.
 Jet-jet interaction gives a contribution which is comparable
 to that from initial fusion. Once  again the decrease at lower
  $p_{T}$ occurs due to the exclusion of jets having $p_{T} <$ 2 GeV.
We also note that
for  $p_{T} \sim$ 8 GeV, the jet-jet interaction contributes at a level
of 10$\%$ of the
 initial fusion. The jet-thermal contribution on the other hand starts
 at about 20 $\%$ of the initial fusion contribution at the lowest $p_{T}$
and rises to about one-third of initial contribution for larger $p_{T}$.
The thermal and the free-streaming contributions are seen to be quite
 small for larger $p_{T}$, though around $p_{T}$ $\approx$ 0 GeV,
they contribute at a level of 10$\%$ of the initial fusion.

The thermal and the free-streaming contributions shown in Figs.~\ref{fig3}
and~\ref{fig4} deserve more attention.
 We see that the two contributions for larger
initial temperatures at LHC differ by a factor of  about 2 or more,
while for the
smaller initial temperature at  RHIC they are of similar
magnitude. This we feel has its origin in the large initial temperature
 which enhances the phase-space contribution to the thermal production
 of charm quarks. In fact in Figs.~\ref{fig5} and~\ref{fig6}, where we plot the
 contributions for the bottom quarks production at LHC, this
 difference again shows up.

Coming back to our earlier observation about the contribution of these
 processes to the back-to-back correlation of charm quarks, we realize
 that the large contribution from processes like jet-jet, jet-thermal
 and even thermal production of charm quarks at LHC is likely to
drastically alter the conclusions about this which were arrived
at from studies~\cite{nuxu1,nuxu2} which
invoked only the initial fusion processes.

As indicated earlier, we give our results (see Figs.~\ref{fig5}
and~\ref{fig6}) for the production of
 bottom quarks at LHC for two initial conditions discussed earlier.
 Of-course we find that the thermal production of bottom quarks is
 quite negligible at LHC. The jet-jet contribution is seen to be at
 a level of 2--6$\%$ of the initial fusion. The jet-thermal contribution
is also negligible.

Thus we feel that a high statistics data may still be able to
discern the contribution of processes other than initial fusion
 to the back-to-back correlation. We shall report this in our
 future publications.

\section{Summary and Conclusions}
We have reported results on charm production at RHIC and LHC and
 bottom production at LHC from initial fusion and multiple scattering
 processes treated as jet-jet interaction and thermal production and
 passage of jets through QGP. Two different initial conditions,
one with early thermalization and other with thermalization
at $\tau_{i} \approx$ 0.5 fm/$c$ have been used. Substantial
production of charm, specially at LHC,
 is seen in addition to the production due to initial fusion.

The jet-jet contribution to charm quarks at LHC even exceeds the
initial fusion contribution for intermediate $p_{T}$ for suitable
 initial conditions. We have argued that since back-to-back correlation
for charm quarks from the processes under consideration could be
 different from those from initial fusion, the results for that,
specially at LHC would be considerably affected. This may even be
 discernible for back-to-back correlation of bottom quarks at LHC.
 As indicated earlier, these results will be published shortly.

One could think of several improvements. The obvious one would be to
include energy loss suffered by the interacting gluons and quarks
before fusion and those by the heavy quarks after production.
 We feel that the inclusion of energy loss before fusion may
have smaller effect on over-all production of the heavy quarks
as these would be limited to the earliest times when the momenta
and the temperature are still very large. The energy loss suffered
by the heavy quarks after the production can not be neglected (see e.g.,
Ref.~\cite{jamil}). This will alter the $p_T$ distribution of the
heavy quarks. However,
as it will affect the heavy  quarks produced by all the processes,
their relative contributions will remain largely unaffected.
Relaxing the condition of chemically equilibrated plasma~\cite{lmw}
will also be useful.
It should be of interest to, alternatively, estimate the prompt charm
 production using  the
colour glass condensate model which predicts different scalings for
different rapidities~\cite{dima22}.

A more complete calculation using parton cascade model ~\cite{bms}
will be reported shortly (see also, Ref.~\cite{zxu}).

We conclude that production of charm quarks at
 LHC and even at RHIC from processes other than initial fusion can be
 large and can play a significant role in our study of back-to-back
correlation. This will have important implications for the study of
the nuclear modification factor $R_{AA}$ as well as large mass
dileptons having their origin in the correlated charm decay~\cite{ramona,shor}.

\section*{Acknowledgments} We thank Dr. Umme Jamil for valuable discussions.
One of us (MY) acknowledges financial support of the Department of Atomic Energy
during the course of these studies.


\begin{thebibliography}{99}
\bibitem{comment} The quark annihilation term in Ref.~\cite{lmw} is
given as proportional to $g_q^2N_f^2$ in present notation. It should
be $g_q^2N_f$ as in Ref.~\cite{rm}, which is also used here. However,
as the quark contribution is very meagre, this has only a marginal
effect on the final results.
\end{thebibliography}
\end{document}